\documentclass[aps,prl,twocolumn,showpacs,superscriptaddress]{revtex4}
\usepackage{graphicx}
\usepackage{amsmath}
\begin{document}

\title{Incomplete fusion versus breakup competition with weakly bound projectiles}
\author{F. A. Souza}
\author{M. G. Munhoz}
\author{J. Takahashi}
\author{R. Liguori Neto}
\author{M. M. de Moura}
\author{A. A. P. Suaide}
\author{E. M. Szanto}
\author{A. Szanto de Toledo}
\affiliation{Instituto de F\'{i}sica - Universidade de S\~ao Paulo, Departamento de
F\'{i}sica Nuclear, C.P. 66318, 05315-970, S\~ao Paulo - SP, Brazil}
\author{C. Beck}
\affiliation{IReS and Universit\'e Louis Pasteur, F-67037 Strasbourg, France}
\author{N. Keeley}
\affiliation{CEA-Saclay DAPNIA/SPhN, France}
\author{N. Carlin}
\affiliation{Instituto de F\'{i}sica - Universidade de S\~ao Paulo, Departamento de
F\'{i}sica Nuclear, C.P. 66318, 05315-970, S\~ao Paulo - SP, Brazil}
\date{\today}

\begin{abstract}
The importance of the breakup channel in the vicinity of the Coulomb barrier
($1<E_{c.m.}/V_{b}<2$) is investigated for the medium weight $^{6}$Li + $^{59}$Co
system. Three-body final-state analysis of light-particle coincident data was carried
out to disentangle, for the first time, the breakup contributions from other
competiting mechanisms. $\alpha-d$ angular correlations show incomplete fusion
components as significant as that from breakup process. Their strong coupling
to total fusion is discussed within a comparison with predictions of continuum
discretized coupled-channel calculations.
\end{abstract}


\maketitle
The study of fusion reactions in the vicinity of the Coulomb barrier provides a
fascinating challenge for theories of quantum tunneling leading to an irreversible
complete fusion (CF) of the interacting nuclei into the compound nucleus (CN)
\cite{Balantekin98,Dasgupta98}. The fusion probability is sensitive to the internal
structure of the interacting ions as well as to influence of the other competing
mechanisms such as nucleon transfer and/or breakup (BU) which are known to affect the
fusion. The fusion cross section enhancement generally observed at sub-barrier
energies is understood in terms of dynamical processes arising from couplings to
collective inelastic excitations of the target and/or projectile. However, in the
case of reactions where at least one of the colliding ions has a sufficient low
binding energy so that BU becomes an important process, conflicting experimental
\cite{Tripathi02,Bychowski04,Signorini03} and theoretical results are reported
\cite{Hussein92,Dasso94,Hagino00,Mackintosh04}.

A great experimental effort, involving both (loosely bound) stable and unstable
nuclei, has been devoted to investigate the specific role of the BU channel
\cite{Dasgupta98,Takahashi97,Badran01}. The weak binding of these systems can also
lead to incomplete fusion (ICF)/transfer (TR) processes playing an important role.
Several attempts to clearly identify ICF in $^{6}$Li and $^{9}$Be induced fusion with
fissile targets \cite{Dasgupta99,Dasgupta02} and medium-mass targets
\cite{Woolliscroft03,PGomes05} have been made. CF requires the formation of CN
containing all the nucleons of both projectile and target. If only part of the
projectile fuses with the target, with remaining fragment emerging from the
interaction region, then ICF is defined  (in this case, the BU process is followed by
fusion) \cite{DiazTorres03}. A tranfer-reemission process may lead to a final state
similar to ICF \cite{Carlin89}.

The recent availability of light-mass radioactive ion beams such as $^{6}$He
\cite{Kolata98,Navin04,DiPietro04,Raabe04}, $^{11}$Be \cite{Signorini04}, and
$^{17}$F \cite{Rehm98}, and the renewed interest on reactions involved in
astrophysical processes \cite{Smith01}, motivated the investigation of fusion
reactions involving very weakly bound and/or halo projectiles around and below the
Coulomb barrier. Clearly a full understanding of the BU process and its effects on
near-barrier fusion is fundamental in order to be able to understand the dynamics of
reactions involving radioative nuclei. This requires systematic and exclusive
measurements covering a wide range of processes, systems and energies. We choose to
study both the total fusion \cite{Beck03} and BU \cite{FSouza04} of $^{6,7}$Li with
the intermediate-mass target $^{59}$Co.

In this Letter we address the competition between the several reaction processes and
CF. A three-body kinematics analysis, in which we separate the contribution of the BU
and the ICF, is presented for the first time.

The experiments were performed at the University of S\~ao Paulo Physics Institute.
The $^{6}$Li beam was delivered by the 8UD Pelletron accelerator with energies
E$_{lab}$ = 18, 22, 26 and 30 MeV, and bombarded a 2.2 mg/cm$^{2}$ thick $^{59}$Co
target. The detection system consisted of a set of 11 triple telescopes
\cite{MMoura01} separated by 10$^{o}$, for which light particles can be detected with
a very low-energy threshold (0.2 MeV for $d$ and 0.4 MeV for $\alpha$ particles).

In this work we will concentrate on $\alpha-d$ coincidences, which are usually fully
attributed to a ``BU process'' \cite{Signorini03}. Depending on the angular
combination, we observe well defined peaks, in the total kinetic energy spectrum,
corresponding to the sequential BU of $^{6}$Li in its first excited state with
E$^{*}$ = 2.19 MeV. No other discrete excited states are observed. When analysing the
$\alpha+d$ coincidence yields from the $^{6}$Li induced reaction, we have to consider
the contributions of other processes than BU, leading to the same particles in the
final state. The processes to be considered are:
\\

$i$) $^{6}$Li + $^{59}$Co $\rightarrow$ $^{6}$Li$^{*}$ + $^{59}$Co $\rightarrow$
$\alpha$ + $d$ + $^{59}$Co

$ii$) $^{6}$Li + $^{59}$Co $\rightarrow$ $\alpha$ + $^{61}$Ni$^{*}$ $\rightarrow$
$\alpha$ + $d$ + $^{59}$Co

$iii$) $^{6}$Li + $^{59}$Co $\rightarrow$ $d$ + $^{63}$Cu$^{*}$ $\rightarrow$
$\alpha$ + $d$ + $^{59}$Co

$iv$) $^{6}$Li + $^{59}$Co $\rightarrow$ $^{65}$Zn$^{*}$ $\rightarrow$ $\alpha$ +
$d$ + $^{59}$Co
\\

Process $i$) is identified as the sequential BU of $^{6}$Li. The final state can also
be reached through a direct BU. Process $ii$) can be identified as incomplete fusion
of $d$ + $^{59}$Co ($d$ ICF) with the subsequent reemission of a deuteron from the
excited $^{61}$Ni. This process could also be considered as a $d$ transfer followed
by a $d$ reemission from the $^{61}$Ni nucleus. The same observations are valid
regarding process $iii$), for which either incomplete fusion of $\alpha$ + $^{59}$Co
($\alpha$ ICF) or $\alpha$ transfer could occur and an $\alpha$ particle is
reemitted. Process $iv$) corresponds to the $\alpha-d$ sequential decay of the
$^{65}$Zn CN. The contribution of CN decay is considered negligible, as confirmed by
predictions from statistical model codes.

In order to investigate the competition of the above processes, we performed a
complete three-body kinematics analysis. As we know the detection angle and energies
of both the $d$ and the $\alpha$ particle, from the three-body kinematics equations
we can determine the energy and emission angle of the remaining $^{59}$Co nucleus and
from this, all the quantities of interest, such as $Q$-values and relative energies.
By generating $Q$-value spectra, we observe from the data, for most of the events in
the final state $\alpha$ + $d$ + $^{59}$Co, the residual $^{59}$Co nucleus mostly in
its ground state, in which the events were gated. Products from the sequential
$^{6}$Li BU $\alpha+d$, are focused inside an angular cone. If we assume that process
$i$) is occurring, the relative energy E$_{\alpha-d}$ between $\alpha+d$ and $d$ in
the rest frame of $^{6}$Li can be calculated. By fixing for instance, the detection
angle of the $\alpha+d$ particle and varying the $d$ detection angle, we can follow
the behavior of the relative energy E$_{\alpha-d}$. In Fig. 1(a) is shown, for
E$_{lab}$ = 29.6 MeV (corrected for the energy loss to the center of the target), the
relative energy E$_{\alpha-d}$ as a function of the deuteron detection angle
$\theta_{d}$, for a fixed angle $\theta_{\alpha}$ = 45$^{o}$ of the $\alpha$
particle. It is interesting to notice that, within the angular range where the
sequential BU of $^{6}$Li in its first excited state is kinematically allowed
(delimited by the two vertical dashed lines), the relative energy E$_{\alpha-d}$ is
constant. The E$_{\alpha-d}$ value is consistent with the sequential BU of $^{6}$Li
in its first excited state. Outside this region it is no longer constant, which
suggests the presence of other processes. In addition to ICF or TR, a direct BU
occuring close to target with strong nuclear field could also give values where the
quantity E$_{\alpha-d}$ is not constant.

\begin{figure}
\begin{center}
\includegraphics[width=1.0\columnwidth]{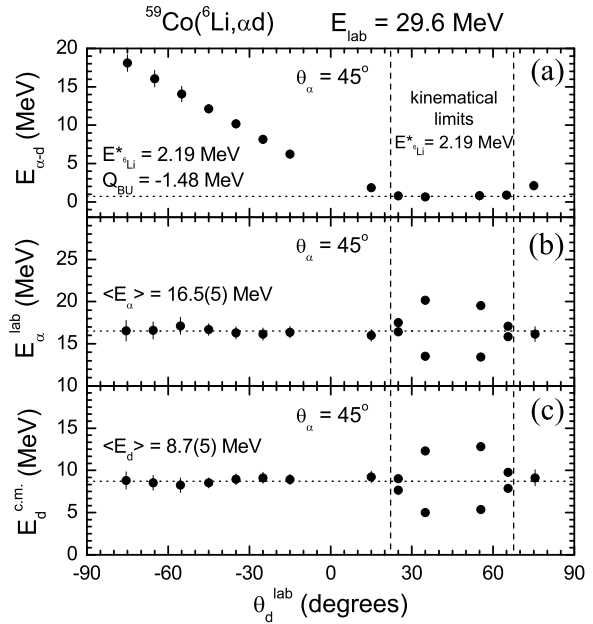}
\caption{(a) Relative energy E$_{\alpha-d}$ as a function of the detection angle
$\theta_{d}$. (b) Average $\alpha$ energy E$_{\alpha}$ as a function of the deuteron
detection angle $\theta_{d}$. (c) $d$ energy in the rest frame of the decaying
$^{61}$Ni as a function of $\theta_{d}$. The angles with two experimental points
correspond to two kinematical solutions for the sequential BU.}
\end{center}
\end{figure}

In Fig. 1(b) we show the behavior of the laboratory $\alpha$ particle kinetic energy
E$_{\alpha}$ as a function of $\theta_{d}$ for a fixed angle $\theta_{\alpha}$ =
45$^{o}$, and for E$_{lab}$ = 29.6 MeV. Here, E$_{\alpha}$ is taken as the centroid
of the experimental coincidence spectra. We observe that the average energy
E$_{\alpha}$ is constant (independent of the momentum of the deuteron), except in the
angular range where the sequential BU of $^{6}$Li is present. Considering now process
$ii$), if this binary process is occurring with an intermediate stage, for a given
angular combination $\theta_{\alpha}$ and $\theta_{d}$ the energy E$_{\alpha}$ is
uniquely determined once the $^{61}$Ni excitation energy is defined. We can then
conclude from Fig. 1(b) that if process $ii$) is dominant over the angular range
where E$_{\alpha}$ is constant, the average $^{61}$Ni excitation energy is also
constant, and in this case E$^{*}_{^{61}Ni}$ = 25 MeV. This is a very important
result as a clear indication of a CN-type reaction. This constant value is consistent
with the assumption of an ICF $d$ + $^{59}$Co where the $d$ has the projectile
velocity.  As a consistency check, we also calculated the $d$ energy in the rest
frame of the decaying $^{61}$Ni. If the $^{61}$Ni is in equilibrium and the
excitation energy is constant over the mentioned angular interval, the $d$ energy
should be constant and independent of the $d$ emission angle $\theta_{d}$. This is
indeed observed, and is shown in Fig. 1(c). The same behavior is observed for other
configurations with different fixed angles $\theta_{\alpha}$. On the other hand, if a
TR is assumed, an optimum $Q$-value can be calculated and in this case the residual
nucleus $^{61}$Ni would have an excitation energy E$^{*}_{^{61}Ni}$ = 22 MeV. This
value is also close to the experimental one. Taking into account the different
relations available for the optimum $Q$-value calculation, this could also be closer
to 25 MeV.

The process defined in $iii$) could also be present in the coincidence data. In order
to check that, the same analysis described above was performed for the situations
with a fixed angle $\theta_{d}$ as a function of $\theta_{\alpha}$. A similar
behavior is obtained, and this leads us to conclude that, within our sensitivity,
there is also a significant contribution from process $iii$). 

\begin{figure}
\begin{center}
\includegraphics[width=1.0\columnwidth]{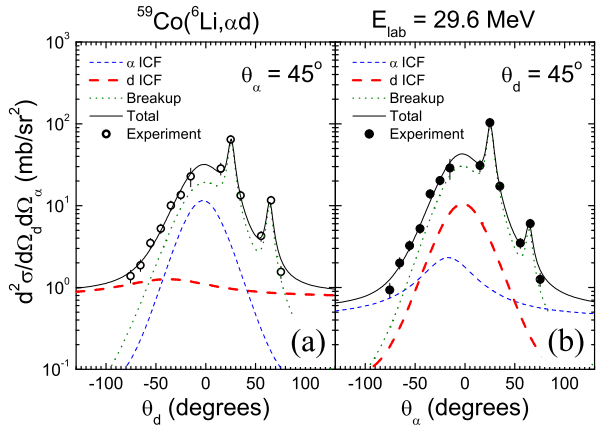}
\caption{(Color online) Experimental in-plane angular correlations and the
predictions for the contribution of $d$ ICF and $\alpha$ ICF. The contribution of the
BU is also shown. The total sum is shown as a solid line.}
\end{center}
\end{figure}

In order to quantify the contribution from processes $i$), $ii$) and $iii$), we
constructed the $\alpha-d$ angular correlation functions through calculation of the
double differential cross sections
${\rm d}^{2}\sigma/({\rm d}\Omega_{\alpha}{\rm d}\Omega_{d})$.
The absolute cross sections, the product of the number of particles in the target per
unit area and number of particles in the beam ($N_{A}N_{B}$) for each run, was
calculated and normalized to the elastic scattering data we measured. The
uncertainties in the experimental points (about 10\% to 40\%) are due to statistics,
the determination of $N_{A}N_{B}$ and, the geometrical determination of the detector
solid angles.

As mentioned above, a priori, events from TR and ICF process are indistinguishable
due to the fact that the intermediate nucleus, in both cases, is populated in the
continuum, and a statistical description for these processes seems adequate
\cite{Oliveira96}. The model we propose as follows is based on an ICF picture.
 
Assuming that for the $d$ ICF ($\alpha$ ICF) process, the excited $^{61}$Ni
($^{63}$Cu) is in equilibrium prior to the $d$ ($\alpha$) reemission, a model can be
utilized to describe the decay. We choose here the approach developed by Halpern
\cite{Halpern59}. It consists of a classical model for emission from a spherical
rotating nucleus. If a rotating nucleus is assumed to rotate around an axis that is
perpendicular to the reaction plane, then the angular distribution of the evaporated
particles is isotropic in the equatorial plane. Due to the centrifugal force, the
yield is concentrated in the equatorial plane and decreases toward the poles. If the
rotating axis is normal to the reaction plane, then the equatorial plane is the
reaction plane.

The yield of evaporation particles as a function of the polar angle $\psi$ defined
with respect to the rotation axis is given by:
\begin{equation}
Y(\psi)=Y_{0}\text{exp}(Xsin^{2}\psi) \label{1}
\end{equation} 
where $Y_{0}$ is a normalization factor and $X$ is the ratio of rotational kinetic
energy to the thermal nuclear energy: 
\begin{equation}
X=0.5(J+\frac{1}{2})^{2}/2IT \label{2}
\end{equation} 
where $J$ is the spin of the rotating nucleus, $I = \mu$$R^{2}$ is the moment of
inertia, and the temperature $T$ can be estimated from $E^{*} = aT^{2}$, with $a$
being the level density parameter.

The angle between the rotational axis and the z axis which is normal to the reaction
plane, is defined to be $\gamma$. The angle $\gamma$ is assumed to be gaussian
distributed. From Eq. (1), the angular distribution is:
\begin{equation}
W(\theta,\phi)=\int\text{d}\gamma\text{ exp}(-\gamma^{2}/2\gamma^{2}_{0})Y(\psi)
\label{3}
\end{equation}
with $\text{cos}\psi=\text{cos}\gamma\text{cos}\phi+\text{sin}\gamma\text{sin}\phi
\text{sin}\theta$.

The angular distribution $W(\theta,\phi)$ is the same as the experimental quantity
${\rm d}^{2}\sigma/({\rm d}\Omega_{\alpha}{\rm d}\Omega_{d})$ in the rest frame of
the rotating nucleus. In the laboratory reference frame, $W(\theta,\phi)$ is centered
at the recoil direction in the primary process. It is important to remark that by
using this model, the angular correlation $\phi$ dependence is also taken into
account. Therefore, the integration is performed in a better way than assuming an
isotropic $\phi$ dependence.

For $d$ and $\alpha$ emission, the best values for $Y_{0}$, $X$ and $\gamma_{0}$ are
obtained from $\chi^{2}$ fits to the angular distributions provided by the
statistical code STATIS \cite{Stokstad72,Carlin89}. For the calculations, a fusion
process $d$ + $^{59}$Co $\rightarrow$ $^{61}$Ni$^{*}$ or $\alpha$ + $^{59}$Co
$\rightarrow$ $^{63}$Cu$^{*}$ was assumed, with a bombarding energy forming the
excited CN in an excitation energy corresponding to the most probable value observed
experimentally.

In Fig. 2(a) we show the $\alpha-d$ angular correlation for a fixed $\theta_{\alpha}$
= 45$^{o}$ and E$_{lab}$ = 29.6 MeV, together with the model predictions for the $d$
ICF and $\alpha$ ICF. The shape of the $\alpha$ ICF correlation is obtained from the
model predictions for the angular correlations with $\theta_{d}$ fixed. In the same
manner, the $d$ ICF shape presented in Fig. 2(b) is obtained from the predictions for
the angular correlations with $\theta_{\alpha}$ fixed. The sequential and direct BU
contribution is also shown in Fig. 2. It is obtained through the subtraction of the
incoherent sum of the $d$ and $\alpha$ ICF contributions, from what would be the best
fit to the data. The two peaks lying around 45$^{o}$ correspond to the sequential BU
of the 2.19 MeV unbound state of $^{6}$Li.

By numerically integrating the angular correlation function in $\theta$ and $\phi$,
for each process, the differential cross section
${\rm d}\sigma/{\rm d}\Omega_{\alpha}$ (${\rm d}\sigma/{\rm d}\Omega_{d}$) is
obtained for the fixed $\theta_{\alpha}$ ($\theta_{d}$). The procedure is repeated
for all the other angular correlations with different fixed $\theta_{\alpha}$ or
$\theta_{d}$.

The BU cross section is obtained with the utilization of the $d$ singles cross
sections. We assume that the $d$ singles spectra present predominantly the
contribution of BU, $d$ ICF and $\alpha$ ICF. The total BU cross section for a given
bombarding energy is then obtained by the difference between the total $d$ singles
cross section and the total $d$ ICF + $\alpha$ ICF cross sections.

\begin{figure}
\begin{center}
\includegraphics[width=1.0\columnwidth]{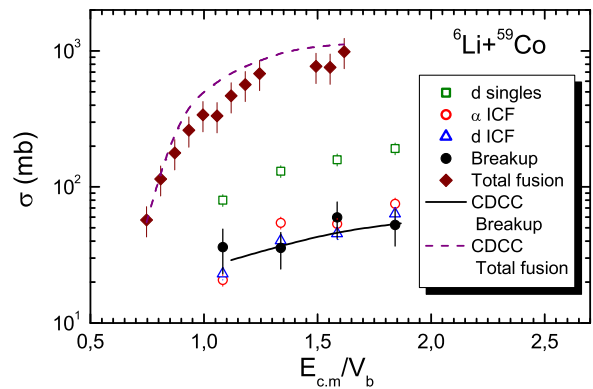}
\caption{(Color online) $^{6}$Li + $^{59}$Co excitation functions for d singles, $d$
ICF, $\alpha$ ICF, BU and fusion \cite{Beck03}. The BU CDCC prediction is shown as a
solid line. The total fusion CDCC prediction \cite{DiazTorres03} is shown as a dashed
line.}
\end{center}
\end{figure}

The exclusive BU cross sections for the resonant states in $^{6}$Li have been
calculated by the CDCC formalism \cite{DiazTorres03} using a cluster folding model
with potentials that describe well the measured elastic scattering angular
distributions \cite{FSouza04}. The CDCC calculations for $^{6}$Li were performed with
the code FRESCO assuming an $\alpha+d$ cluster structure, similar to that described
in \cite{Keeley03}. The binding potentials between $\alpha+d$ were taken from
\cite{Kubo72} and the $\alpha+d$ continuum was discretized into series of momentum
bins of width $\delta$$k$ = 0.2 fm$^{-1}$ (up to $k$ = 0.8 fm$^{-1}$) for $L$ = 0,1,2
for $^{6}$Li, where $\hbar$$k$ denotes the momentum of the $\alpha+d$ relative
motion. All couplings, including continuum-continuum couplings, up to multipolarity
$\lambda$ = 2 were incorporated \cite{Keeley02a}. The total calculated BU
cross-sections for $^{6}$Li were obtained by integrating contributions from states in
the continuum up to 11 MeV. The results of full CDCC calculations are displayed in
Fig. 3. The solid line corresponds to the final results of the BU CDCC calculations,
and the symbols represent the experimental excitation functions obtained for the $d$ singles,
and for the $d$ ICF, $\alpha$ ICF and the BU processes. The $d$ ICF and $\alpha$ ICF
cross sections are comparable to the BU cross sections for all the bombarding energies.

In summary, we have presented a three-body kinematics analysis of coincidence
measurements with the aim of disentangling the contribution of BU and competing
processes in reactions with weakly bound nuclei. From the analysis of $\alpha-d$
coincidences of $^{6}$Li + $^{59}$Co at several near-barrier energies, we observed a
significant contribution from the ICF process, with a cross section comparable to the
one from BU process. This suggests that ICF should be accounted for in the coupled
channels calculations to explain the fusion of weakly bound nuclei inhibition and/or
enhancement at near-barrier energies.

\begin{acknowledgments}
The authors thank FAPESP and CNPq for the financial support and also Prof. C.
Signorini and Prof. B. R. Fulton for the careful reading of the manuscript.
\end{acknowledgments}

\end{document}